\documentclass[twocolumn,preprintnumbers,showpacs,showkeys,amsmath,amssymb,aps]{revtex4}
\usepackage{graphicx}
\usepackage{dcolumn}
\usepackage{bm}
\usepackage{epsf}

\newcommand{\edc}{\end{document}}
\newcommand{\bb} {\begin{thebibliography}}
\newcommand{\eb}{\end{thebibliography}}
\newcommand{\bi}[1]{\bibitem{#1}}
\newcommand{\bc}{\begin{center}}
\newcommand{\ec}{\end{center}}

\newcommand{\be}{\begin{equation}\small}
\newcommand{\ee}{\end{equation}\normalsize}
\newcommand{\bea}{\begin{eqnarray}}
\newcommand{\eea}{\end{eqnarray}}
\newcommand{\ba}{\begin{array}{l}   }

\newcommand{\ea}{\end{array}}

\newcommand{{\vergul}}{  ,}

\begin{document}
\preprint{2011-}
\title{Generalized Eaton Lens at Arbitrary Refraction Angles}
\author{Sang-Hoon  \surname{Kim}} \email{shkim@mmu.ac.kr }
\affiliation{
Division of Liberal Arts $\&$ Sciences,
Mokpo National Maritime University,
Mokpo 530-729, R. O. Korea}
\date{\today}
\begin{abstract}
We extended the refraction angle of the Eaton lens into arbitrary
angles. The refractive index of the Eaton lens is not analytical and
can be obtained by numerical calculations only  except in the case of the
 retroreflector. We introduced an approximation for the refractive index of the
generalized Eaton lens. It is simple but very close to the exact
values obtained from numerical calculations.
\end{abstract}
\maketitle


Trajectories of light can be controlled by prisms or a
combination of mirrors. At the same time it can be achieved by
controlling the refractive index(RI) of a lens. It is a
GRIN(Gradient Index) lens. Once it was thought to be
unrealistic or very difficult to realize, but the recent development of
transformation optics and high RI materials from metamaterial
techniques have opened a new way to control the trajectories of waves.

The Eaton lens is a typical GRIN lens where the RI varies from one to
infinity.
It has a singularity at the center of the lens that the RI goes to
infinity and it originates from a singularity of dielectrics.
The speed of light reduces to zero at the singularity, therefore, the
lens can change the trajectories of waves into any direction.

For the three specific refraction angles such as
$90^o$(right-bender), $180^o$(retroreflector), and
$360^o$(time-delayer), the Eaton lens has been studied recently
\cite{hannay,tomas,danner}.
The RI  of the Eaton lens is given as a function of radius, but
it is not analytic except for the special
 refraction angle of  the retroreflector.
It is obtained by numerical calculations only. Then that makes the
application of the lens to be inconvenient and is not easy to handle.

In this short note, the RI of Eaton lens is extended to arbitrary
refraction angles. The generalized form of the RI is not analytic.
To overcome the inconvenience, an approximated form of the RI is
suggested for easy and practical use. It comes from a linear
approximation of the retroreflector. Needless to say it should be
very close to the numerical values at some realistic refraction
angle ranges.


The RI for the three specific refraction angles have been studied
already. For symmetric and spherical lenses they are known as
\cite{danner}
 \bea n^2 &=&
\frac{1}{nr}+\sqrt{\frac{1}{n^2 r^2}-1} \hspace{10pt} (\theta =
90^o),
\label{10} \\
 n &=& \sqrt{\frac{2}{r}-1} \hspace{52pt} (\theta = 180^o),
\label{12} \\
 \sqrt{n} &=& \frac{1}{nr}+\sqrt{\frac{1}{n^2 r^2}-1} \hspace{9pt}
 (\theta = 360^o).
\label{14} \eea $n=1$ for $r \ge 1$, where  $n$ is the relative RI.
 The $r$ is the radial position between 0 and 1, and
the actual radial position is given by $a r$, where $a$ is the
radius of the lens.

\begin{figure}
\resizebox{!}{0.11\textheight}{\includegraphics{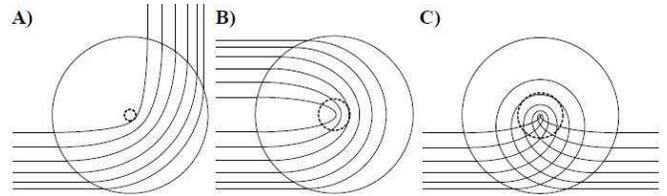}}
\caption{\label{fig:eaton1} The trajectories of Eaton lenses at
three specific angles from Ref. \cite{danner}.
(A)$90^o$, (B)$180^o$, (C)$360^o$.}
\end{figure}

The importance of the above three lenses has been discussed.
 The right-bender in Eq. (\ref{10})  has been studied
 in a bending surface plasmon polaritons \cite{zent}.
 The retroreflector in Eq. (\ref{12}) is a device
 that returns the incident wave back to their source \cite{ma}.
The time-delayer in Eq. (\ref{14}) is a time clocking device that
the incident wave leaves in the direction of incidence as if the
lens were not present. The trajectories for the above three cases
has been plotted by Danner and Leonhardt \cite{danner}. It is shown
in Fig.~\ref{fig:eaton1}.

The RI of arbitrary refraction angles is easily derived from Ref.
\cite{hannay}.  RI is identical to particle trajectories of equal
total energy $E$ in a central potential $U(r)$.
From the conservation of mechanical energy
inside and outside the lens, the kinetic energy of a particle in the
medium of RI $n$
 is written as $(1/2)m n^2 v^2 = E-U$, where $m$ is the mass
  and $v$ is the velocity of the particle inside the lens.
Then, $\int n ds$ or $\int \sqrt{E-U}ds$ should be stationary by Fermat's principle.
The potential of the Eaton lens corresponds to $U \propto -1/r$
\cite{hannay}. On the other hand the potential Luneburg lens
corresponds to $U \propto r^2$ \cite{morgan}. The trajectory of the
Eaton lens is an analogue of Kepler' scattering problem \cite{landau}.

\begin{figure}
\resizebox{!}{0.22\textheight}{\includegraphics{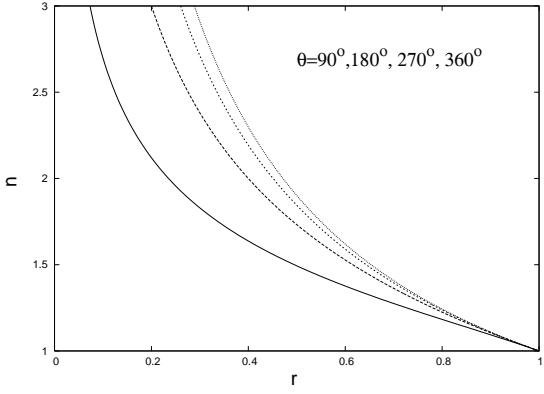}}
\caption{\label{fig:eaton2} The refractive indexes at $\theta=90^o,
180^o, 270^o$, and $360^o$ from left to right. }
\end{figure}

Replacing $\chi +\pi=\theta_{final}-\theta_{initial}$ into $\theta$
in Ref. \cite{hannay},
we obtain a generalized form of the RI of
Eaton lens at arbitrary refraction angles as
\be
 n^{\pi/\theta} = \frac{1}{nr}+\sqrt{\frac{1}{n^2 r^2}-1},
\label{20} \ee
where $\theta$ is any radian angle and
 $n=1$ for $r \ge 1$.
The refraction angle can be generalized as $\theta=(2N+1/2)\pi$ for
 right-bender, $\theta=(2N+1)\pi$ for  retroreflector, and
 $\theta=2N\pi$ for  time-delayer. Note that $N$ is an integer.
The RI is not analytic still. We calculated numerically and plotted
 RI for four specific angles  in Fig.~\ref{fig:eaton2}

\begin{figure}
\resizebox{!}{0.22\textheight}{\includegraphics{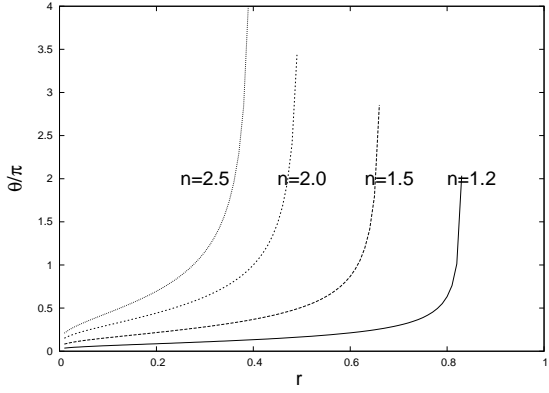}}
\caption{\label{fig:eaton3} The refraction angles at various
refractive index and radius.}
\end{figure}

Taking logarithm we have a more convenient form of the relations
between RI and the radius.
\be
\frac{\theta(n,r)}{\pi} = \frac{\log
n} {\log \left( \frac{1}{nr}+\sqrt{\frac{1}{n^2 r^2}-1} \right)}.
\label{30}
\ee
 For any given radius and RI,  we can predict the
refraction angle from Eq. (\ref{30}).
It is plotted in Fig.~\ref{fig:eaton3}.
Large RI changes the
trajectories at small radius. They are almost inversely proportional
to each other.

\begin{figure}
\resizebox{!}{0.22\textheight}{\includegraphics{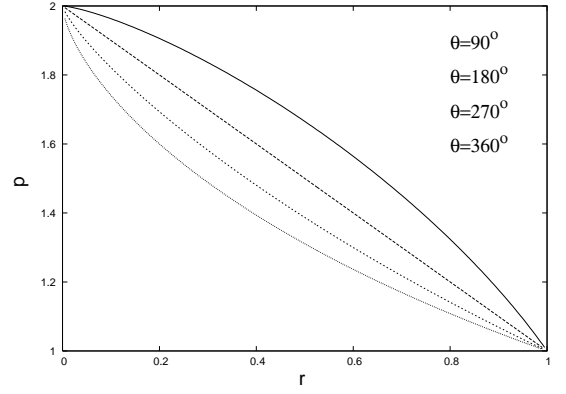}}
\caption{\label{fig:eaton4} $p(r)$ as a function of radius.
$\theta=90^o, 180^o, 270^o$, and $360^o$ from top to bottom. $p(r)$
is linear for $\theta=180^o$}
\end{figure}

Eq. (\ref{30}) is relatively convenient to find
the relations between RI and radius
 than Eq. \ref{20},
but still not  practical. We need an analytic form in lab to apply.

From Eq. (\ref{20}) we obtain $n(r)$ at two boundaries. As $r \sim
0$,  then $n \sim (2/r)^{\theta/(\pi + \theta)}$. And as $r \sim 1$,
then $n \sim (1/r)^{\theta/(\pi + \theta)}$. Therefore, we can write
$n(r)$ at the range of $0< r \le 1$ as the following compact form
\begin{equation}
 n(r) = \left\{ \frac{p(r)}{r}\right\}^{\frac{\theta}{\pi+\theta}},
\label{40}
\ee
where $0 < p(r) \le 1$.

The  $p(r)$ is a bounded and analytic function of radius between 1
and 2. It is obtained from \be
 p(r) = r
n(r)^{\frac{\pi+\theta}{\theta}}.
 \label{45}
 \end{equation}
$p(r)$ is plotted in Fig.~\ref{fig:eaton4}. It is monotonically
decreasing from $r=0$ to $r=1$. It is pretty linear. Therefore, we
can take the retroreflector approximation or a linear approximation
as $p(r)=2-r$. Then, the RI has a simple form as
\be
n(r) \simeq
\left( \frac{2}{r} -1\right)^{\frac{\theta}{\pi+\theta}}.
\label{50}
\ee

\begin{figure}
\resizebox{!}{0.22\textheight}{\includegraphics{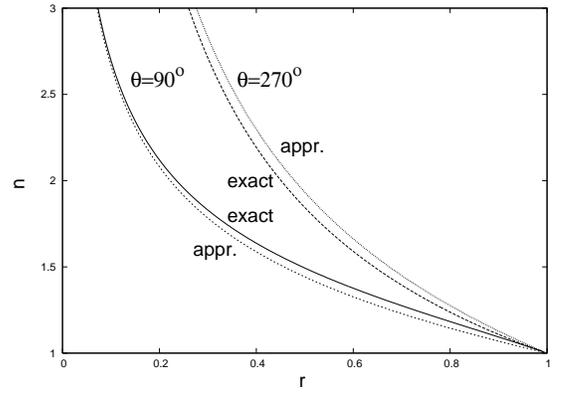}}
\caption{\label{fig:eaton5} Comparison between the exact value
obtained by numerical calcultion and the retroreflector
approximation for $\theta=90^o, 270^o$. They are match exactly for
$\theta=180^o$.
 They matches at both ends of $r \sim 0$ and $r \sim 1$ at every angles. }
\end{figure}

The effectiveness of the approximation is examined by comparison
with some exact values obtained from numerical calculation in
Fig.~\ref{fig:eaton5}. We see the approximation is pretty good at
the range of $0< \theta <2\pi$.

The time-delayer with $\theta=2N\pi$ is an invisible sphere. The
general form with $N$ turn  can be represented easily using $p(r)$
in Eq. (\ref{40}). When it has $N$ turn, the time delay $\triangle
t$ is obtained as
\be
\triangle t = \frac{2N\pi r n}{v_o}
=\frac{2N\pi r }{v_o}\left\{ \frac{ap(r)}{r} \right\}^{\frac{2N}{2N+1}},
\label{55}
\ee
where $v_o$ is the background velocity outside the
lens and $a$ is the radius of the lens. If $N \gg 1$, then $2 N \pi
a/v_o \le \triangle t \le 4 N \pi a/v_o$. Therefore, $a$ and $N$ are
the two main factors that decide the time delay.


The Eaton lens is a typical GRIN lens with a complicated RI which has been
studied at some specific angles. We derived the RI of Eaton lens at
arbitrary refractive angles. It is not analytical at most angles and
inconvenient to find the values easily. We introduced a
retroreflector approximation for the RI. It is simple but very
close to the exact values of the RI that are obtained from numerical
calculation.

We would like to thank C. K. Kim and S. H. Lee for useful discussions.
This research was supported by Basic Science Research Program through
the National Research Foundation of Korea(NRF)
funded by the Ministry of Education, Science and Technology(2011-0009119)

\bb{99}
\bi{hannay} J. H. Hannay and T. M. Haeusser, J. Mod. Opt.  {\bf 40},
1437 (1993).
 \bi{tomas} T. Tyc and U. Leonhardt, N. J. Phys.
{\bf 10}, 115038 (2008).
\bi{danner} A. J. Danner and U. Leonhardt,
2009 Conference on Lasers and Electro-Optics(CLEO), Baltimore, MD,
U. S. A. (2009).
 \bi{landau} L. D. Landau and E. M. Lifshitz, {\it
Mechanics}, 3rd ed. Sec. 18 (Pergamon, Oxford, 1976).
 \bi{zent} T. Zentgraf, Y. Liu, M. H. Mikkelson, J. Valentine, and X. Zhang,
Nature, Nano, {\bf 6}, 151 (2011).
\bi{ma} Y. G. Ma, C. K. Ong,
T. Tyc, and U. Leonhardt, Nature, Mat. {\bf 8}, 639 (2009).
\bi{morgan} S. P. Morgan, J. Appl. Phys. {\bf 29}, 1358 (1958).

\eb
\edc